\documentstyle[twocolumn,aps]{revtex}
\begin{document}

\draft

\input epsf

\title{Thermodynamic Capacity of a Protein}
\author{Thomas M.\ A.\ Fink$ ^{\dag}$ and Robin C.\ Ball$ ^{\S}$}
\address{Theory of Condensed Matter, Cavendish Laboratory, 
Cambridge CB3 0HE, England \\
{\rm $^{\dag}$tmf20@cam.ac.uk} \quad 
{\rm $^{\S}$r.c.ball@warwick.ac.uk} \quad
{\rm $^{\dag}$http://www.tcm.phy.cam.ac.uk/$\,\tilde{}\,$tmf20/} 
}

\maketitle

\begin{abstract}
We show that a protein can be trained to recognise multiple conformations,
analogous to an associative memory, and provide capacity calculations based on
energy fluctuations and information theory.
Unlike the linear capacity of a Hopfield network, the number of conformations
which can be remembered by a protein sequence depends on the size of the
amino acid alphabet as $\ln A$, independent of protein length.
This admits the possibility of certain proteins, such as prions,
evolving to fold to independent stable conformations, as well as novel
possibilities for protein and heteropolymer design.
\end{abstract}

\narrowtext


\input epsf

It is widely thought to be a design feature of real proteins that their native,
biologically active state is both a deep global energy minimum 
and has a funnel of low energy configurations leading toward it \cite{Dill97}.
The deep well ensures that a significant fraction of protein molecules
occupy the native state at any given moment.
The funnel guides the molecule to fold to its stable native conformation 
in a time much less than that required for it to explore all configurations, 
thus avoiding the so-called Levinthal paradox.

Inverse protein folding, or protein design, consists of designing a sequence
of amino acids that stably and quickly folds to a desired target conformation.
This process may be expressed in the context of the energy
landscape, to which each sequence corresponds.
For each compact conformation $\Gamma_{\rm c}$, there are typically a 
myriad of sequences which fold to it \cite{FB97}.  
The set of sequences which fold to $\Gamma_{\rm c}$ corresponds to
those energy landscapes whose global minima lie above the target.
Most of these will possess nominally global (shallow) minima and fold 
in very long rather than biological time scales \cite{Dill97}.
Of those which are deep, and hence thermodynamically stable, 
fewer yet will resemble broadly sloping funnels.
It is this last group of energy landscapes, and hence sequences, to which 
natural proteins are believed to correspond.  Not surprisingly, we wish to 
select for similar features when engineering artificial proteins.

In this sense, protein design corresponds to choosing from the spectrum 
of all possible sequences a sequence whose landscape possesses the 
attributes we desire.
Because the spectrum is finite, however, we are not free to insist on an 
arbitrary topography; some landscapes have wells too deep or too numerous 
to be practicable.

In this Letter we investigate the fundamental limit on the introduction of 
deep (thermodynamically stable) minima into the protein energy landscape
\cite{FinkThesis}.
We estimate the typical maximum depth of the ground state well in a
sequence trained to fold to a unique conformation.
By analogy with the theory of associative neural networks (ANNs) \cite{DA89},
we show how protein design can be generalised to provide recognition of 
several conformations rather than a single target state.  
We find that the number of conformations that a protein can recall is 
limited and calculate its capacity.
Remarkably, the capacity depends not on protein length but on the number of 
amino acid species. 

The ability of a protein sequence to encode multiple conformations has
immediate implications on our understanding of prions and other multi-stable 
proteins.  In his Nobel lecture \cite{SP98}, Prusiner concludes
`The discovery that proteins may have multiple biologically active
conformations may prove no less important than the implications of prions for
diseases.  How many different tertiary structures can [a protein] adopt?
This query not only addresses the issue of the limits of prion diversity but
also applies to proteins as they normally function within the cell\ldots.'
In addition to predicting multi-stable proteins, our results suggest
that artificial heteropolymers may be engineered to fold to 
multiple targets as well.
We discuss possibilities for implementing target control to this end.

Our thermodynamic capacity result --- that the manipulation of the energy
landscape by the introduction of deep minima is limited --- can be
generalised.
We investigate the kinetic capacity of a protein, {\it i.e.}, the 
limit on the size of a folding funnel, in a separate Letter \cite{FBCK00}.


{\bf Proteins as Associative Memories} 
A lattice protein consists of a sequence $S$ of $N$ amino acids, or monomers, 
each of which can take on one of $A$ possible species. 
We denote the species of the $i$th monomer of $S$ by $S_i$, and 
monomers $i$ and $j$ interact according to the $N\times N$
extended pair potential $\tilde{U}$, where $\tilde{U}_{ij} = U_{S_i S_j}$
and $U$ is the $A\times A$ pair potential.

Protein conformations may be represented by the contact matrix $C$, where 
$C_{ij} = 1$ if monomers $i$ and $j$ are nearest neighbours and 0 otherwise.
Contacts between monomers adjacent along the protein chain are preserved
and cannot influence the folding dynamics, so we exclude these from 
the contact map.  For compact conformations, each interior
monomer is surrounded by its chain neighbours plus $z'$ others, where $z'$
(the effective coordination number) is two less than $z$ (the lattice
coordination number).
Contact patterns are thought to be a unique representation of compact 
conformations and we approximate them as independent.

Protein folding may be considered pattern recognition in as much as the
protein rapidly organises itself into the target pattern $C$ upon 
entering the target basin of attraction (funnel).
By analogy with pattern association, this idea may be generalised to
the recognition of multiple patterns.  This raises the question of how to
train the sequence to recognise more than one conformation.
For lattice models, Shakhnovich and co-workers \cite{Shak94,Gutin95} 
have explored the folding of sequences designed to minimise a conformation's 
absolute and relative energies.  The essence of the training technique 
is to embed the protein into the target conformation and optimise stability over
sequence space; the resulting (near-optimal) sequence spontaneous folds to the
target.
The dilute representation of conformations by contact patterns suggests that 
we can superimpose $p$ patterns without saturation \cite{SCALING}, 
providing us with a total pattern to which we train in the usual way.
This is essentially equivalent to the method used to 
select bi-stable 36-mers in \cite{Shak98}.


{\bf Energy Function}
The energy of a sequence in conformation $C$ may be conveniently expressed

\begin{equation}
E = {1\over 2} \sum_{ij=1}^N C_{ij} \tilde{U}_{ij}.
\label{ASSOC_A}
\end{equation}

\noindent For a sequence trained to have minimal energy in conformation $\Gamma_\mu$,
the energy appears as

\begin{equation}
E^{\rm min}_{\mu} = \min_{\tilde{U}} \Big[{1\over 2} \sum_{ij=1}^N C_{\mu_{ij}} \tilde{U}_{ij}\Big]
        = {1\over 2} \sum_{ij=1}^N C_{\mu_{ij}} \tilde{U}^*_{ij},
\label{ASSOC_B}
\end{equation}

\noindent where minimisation is over all $\tilde{U}$ corresponding to valid
sequences and $\tilde{U}^*$ minimises $E_{\mu}$. 
The energy of a fixed sequence $S_\nu$ folded to its ground state conformation is 

\begin{equation}
E^{\rm min}_{\nu} \! \equiv \! E^{\rm min}_{\rm cp}
        = \min_{C}\Big[{1\over 2} \sum_{ij=1}^N C_{ij} \tilde{U}_{\nu_{ij}}\Big]
        = {1\over 2} \sum_{ij=1}^N C_{ij}^* \tilde{U}_{\nu_{ij}}, \! \!
\label{ASSOC_B1}
\end{equation}

\noindent where minimisation is over all $C$ corresponding to valid
conformations and $C^*$ minimises $E_{\nu}$. 
As is common usage, we refer to the quantity $E_\nu$ for an untrained sequence 
as the copolymer energy $E_{\rm cp}$.

Throughout this Letter, the energy of a sequence realised in a particular
conformation is indicated by $E$, 
while the Hamiltonian with which a sequence is trained (generally the linear 
combination of the energies realised in a number of conformations) is denoted 
by $H$.  


{\bf Capacity from Energetics}
We consider the thermodynamic capacity of a protein, that is, the 
number of conformations $p$ that we can train the sequence to make 
simultaneously thermodynamically stable.
For a protein to fold to a single target conformation, it is necessary that 
the energy of the trained sequence realised in that conformation,
$E_{\mu}^{\rm min}$, be below 
the minimum fluctuations of the energy elsewhere,
thereby making the target minimum global.  
Since the trained sequence is not correlated with distant conformations,
energy fluctuations away from the target structure are statistically
equivalent to those of a random copolymer sequence.  We therefore require 
that the trained energy be less than the minimum energy of a
random sequence, that is, $E_{\mu}^{\rm min} < E_{\rm cp}^{\rm min}$.
Folding to a set of $p$ conformations requires that the minimum
energy of all of these lie below $E_{\rm cp}^{\rm min}$.


We first estimate the typical minimum copolymer energy $E_{\rm cp}^{\rm min}$.
Recalling that each row (or column) of the contact map $C$ has $z'$ bonds, 
the quantity $E_{\rm cp}$ from (\ref{ASSOC_B1}) (before minimisation) is 
the sum of ${z'N\over 2}$ bonds. 
Since the extended pair potential $\tilde{U}$ of the copolymer from 
(\ref{ASSOC_B1}) is untrained, these contact energies are uncorrelated and 
may be considered random.  
Assuming a distribution of bonds with zero mean (as is the case of
that found in \cite{MJ96})
and standard deviation $\sigma$, 
we find, in accordance with the central limit theorem, that $E_{\rm cp}$
is distributed as

\begin{equation}
f(E_{\rm cp}) \simeq {1 \over \sqrt{2 \pi} \sigma_{\rm cp}} \exp(-{E_{\rm cp}^2\over 2 \sigma_{\rm cp}^2}),
\label{ASSOC_P}
\end{equation}

\noindent where $\sigma_{\rm cp}^2 = {z'N\over 2} \sigma^2$.
This estimation is valid out to $|E_{\rm cp}|$ of order ${z'N\over 2} \sigma$.
The ground state energy $E_{\rm cp}^{\rm min}$ is the least of
all possible samples of (\ref{ASSOC_P}), each of which corresponds to a unique
conformation.  Since the number of compact conformations of an $N$-mer 
grows as $\kappa^N$, where $\kappa \simeq 1.85$ on a cubic 
lattice \cite{Pande94}, the energy of
the ground state is the minimum of $\kappa^N$ samples of $f(E_{\rm cp})$.


What is the minimum of $M$ samples of a random variable $X$ 
distributed according to a gaussian $g(x)$? 
For convenience we assume zero mean and standard deviation $\sigma_X$.
The probability distribution of $x$ being the minimum of $M$ samples of 
$X$ is given by

\begin{equation}
g^{\rm min}(x) = M g(x) \bigl (1 - G(x)\bigr )^{M-1},
\label{ASSOC_J}
\end{equation}

\noindent where $G(x) = \int_{-\infty}^x g(x'){\rm d}x'$
is the usual cumulative distribution.
Maximising $g^{\rm min}$ with respect to $x$ yields the 
transcendental equation
$x^{\rm min} (1 - G(x^{\rm min})) = - \sigma^2 (M-1) g(x^{\rm min})$,
where $x^{\rm min}$ is the minimum of the $M$ realisations of $X$.
For reasonably large $M$, $G(x)$ is small and we 
estimate $x^{\rm min}$ as

\begin{equation}
x^{\rm min} \simeq -\sqrt{2} \sigma_X \sqrt{\ln M}.
\label{ASSOC_L}
\end{equation}


By way of (\ref{ASSOC_L}), we can express the ground state energy 
$E_{\rm cp}^{\rm min}$ as

\begin{equation}
E_{\rm cp}^{\rm min} \simeq -\sqrt{2} \sigma_{\rm cp} \sqrt{\ln(\kappa^N)}
                  = -\sqrt{z'} N \sigma \sqrt{\ln \kappa}.
\label{ASSOC_Q}
\end{equation}


We now approximate the typical energy of a sequence optimally trained to a set
of $p$ target conformations and arranged in one of these configurations.
The total contact map, to which we train by energy minimisation
with respect to the sequence \cite{Shak94}, 
is defined as a linear superposition of the $p$ corresponding contact maps,
that is

\begin{equation}
C_{{\rm tot}_{ij}} = \sum_{\mu = 1}^p C_{\mu_{ij}}.
\label{ASSOC_C}
\end{equation}

The minimum Hamiltonian associated with the total contact map may 
then be written

\begin{equation}
H_{\rm tot}^{\rm min} = {1\over 2} \sum_{ij = 1}^N C_{\rm tot_{ij}} \tilde{U}_{ij}^*
        = {1\over 2} \sum_{ij=1}^N \sum_{\mu=1}^p C_{\mu_{ij}} \tilde{U}^*_{ij},
\label{ASSOC_D}
\end{equation}

\noindent where here $\tilde{U}^*$ minimises $H_{\rm tot}$.
It is simply the sum of the $p$ individual conformational energies
of the sequence implied by $\tilde{U}^*$.
We re-express the right side of (\ref{ASSOC_D}) as the sum over $i$ of the total 
energy associated with monomer $i$, $H_{{\rm tot}_i}$, each minimised with
respect to the choice of amino acid at monomer $i$,

\begin{equation}
H_{\rm tot}^{\rm min} = \sum_{i=1}^N \min_{S_i} [H_{{\rm tot}_i}];
\label{ASSOC_G}
\end{equation}

\noindent $H_{{\rm tot}_i}$ is obtained by summing over 
the connections to monomer $i$,

\begin{equation}
H_{{\rm tot}_i} = {1\over 2} \sum_{j=1}^N \sum_{\mu=1}^p C_{\mu_{ij}} 
\tilde{U}_{ij}.
\label{ASSOC_H}
\end{equation}

\noindent Since $C$ has $z'$ 
bonds connecting to monomer $i$, each $H_{{\rm tot}_i}$ is 
the sum of ${z'p\over 2}$ random interaction energies freely chosen 
from the pair potential \cite{FRUSTRATION}.
As before, we approximate the distribution of $H_{{\rm tot}_i}$ by its central 
limit theorem form; it is a gaussian with variance
$\sigma_{{\rm tot}_{i}}^2 = {z'p\over 2} \sigma^2$.
This estimation is valid out to $|H_{{\rm tot}_i}|$ of order ${z'p\over 2} \sigma$.

The Hamiltonian 
$H_{{\rm tot}_i}$ at each monomer is minimised with respect to the choice 
of amino acid by choosing the smallest of $A$ samples from the 
distribution of $H_{{\rm tot}_i}$ --- again we wish to estimate the minimum 
of many samples of a gaussian.  By way of (\ref{ASSOC_L}) \cite{CLT},
we find that

\begin{equation}
H_{\rm tot}^{\rm min} \simeq -\sqrt{2} N \sigma_{{\rm tot}_i} \sqrt{\ln A}.
\label{ASSOC_M}
\end{equation}

\noindent When the trained sequence is in one of the $p$ target structures,
the average energy of the sequence is given by

\begin{equation}
E_{\mu}^{\rm min} \simeq {H_{\rm tot}^{\rm min} \over p}
         \simeq -{\textstyle \sqrt{z'\over p}} N \sigma \sqrt{\ln A}.
\label{ASSOC_N}
\end{equation}

\begin{figure}
\begin{center}
  \hspace*{-0.26in}\input{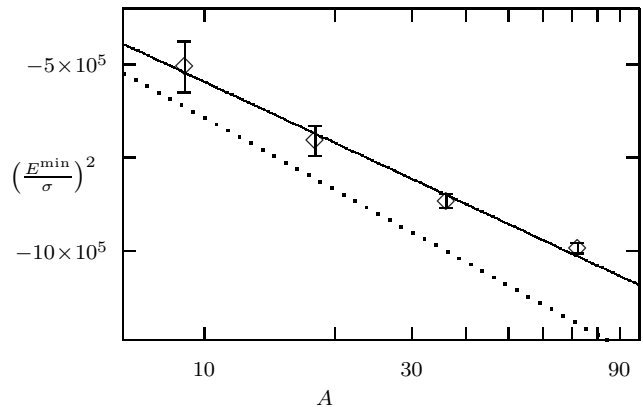} \hspace*{-0.12in}
  \caption{(Negative) square of protein stability ${E^{\rm min}\over \sigma}$ as a 
function of number of amino acid species $A$ (log-linear).
Proteins were trained to fold to a single $6\times 6\times 6$
conformation with periodic boundary conditions by optimisation over sequence
space under constant composition.  
The dotted line was generated by (\ref{ASSOC_N})$\vert_{p=1}$; 
introducing the prefactor $0.847$ gives the solid line.  
Data are shown for $A = 9,18,36$ and $72$ species, for each of which the mean 
and standard deviation were calculated from 12 runs with independent 
random pair potentials.
}
      \label{capacity_ann_approx}
  \end{center}
\end{figure}

Equation (\ref{ASSOC_N}) and results from simulation are plotted in Figure 
\ref{capacity_ann_approx} for $p=1$.
Apart from a prefactor of $0.847$, the predicted dependence of well depth
on $A$ is in good agreement with observation.  
Calculations for $p>1$ are ongoing and will be presented elsewhere.

Comparing the minimum copolymer energy (\ref{ASSOC_Q}) and the minimum 
energy of the trained sequence (\ref{ASSOC_N}) yields

\begin{equation}
\textstyle p_{\rm max} \simeq {\ln A \over \ln \kappa}.
\label{ASSOC_R}
\end{equation}


{\bf Capacity from Information Theory}
The thermodynamic capacity of a protein may also be derived via
information theory.
Consider the transmission of a message, which has been encoded as an $N$ 
letter sequence.
The message is decoded empirically by constructing the protein corresponding to
the sequence (either {\it in vitro} or via computer simulation), allowing 
it to fold 
and observing the $p$ most 
occupied, and consequently lowest, target conformations.  

The information retrieved by learning a single conformation may be
determined as follows.
Given $\kappa^N$ possible compact conformations,
the information contained in one conformation is equivalent to the 
number of bits
necessary to express a number between 1 and $\kappa^N$, 
{\it viz.}, $\ln_2(\kappa^N)$.
Since the $p$ target configurations are assumed to be independent, 
the total retrieved information scales linearly with $p$, that is,
$I_R  = p N \ln_2 \kappa$.

The information transmitted may be similarly determined.
Since the number of sequences grows as $A^N$, the information associated
with a sequence is $\ln_2(A^N)$,
and the total transmitted information \cite{INFO_U} is
$I_T = N \ln_2 A$. 

Information theory dictates that the information retrieved must not be 
greater than the information transmitted, that is,

\begin{equation}
p N \ln_2 \kappa \leq N \ln_2 A. 
\label{ASSOC_S}
\end{equation}

\noindent It readily follows that the bound on $p$ is

\begin{equation}
\textstyle p_{\rm max} = {\ln A \over \ln \kappa},
\label{ASSOC_T}
\end{equation}

\noindent which is identical to the result (\ref{ASSOC_R})
deduced from fluctuations in the energy landscape.


{\bf Discussion of Capacity}
Our bound on capacity has been derived in two ways: by comparison of the 
trained and copolymer minimum energies, which depends on the method of training
(in our case the superposition rule), and by an information 
theoretic argument, which does not.  
The equality of the two results suggests that our constant capacity result 
is not a shortcoming of the superposition rule.

That our bound on memory is independent of chain length $N$ may seem 
surprising given that the capacity of a fully connected ANN grows 
linearly with the number of neurons $n$.
The resolution is that, in both cases, the number of patterns which 
can be stored is of order the number of connections divided by the 
number of nodes.
In the case of a protein the number of active connections (contacts) is 
restricted to of order $N$, whereas for an ANN all $n^2$ connections 
are allowed to contribute significantly. 
The divisor arises because the amount
of information in a pattern is proportional to $N$ and $n$, respectively.

\begin{figure}[t!]
  \begin{center}
    \leavevmode
    \epsfxsize=6.7cm
    \epsfbox{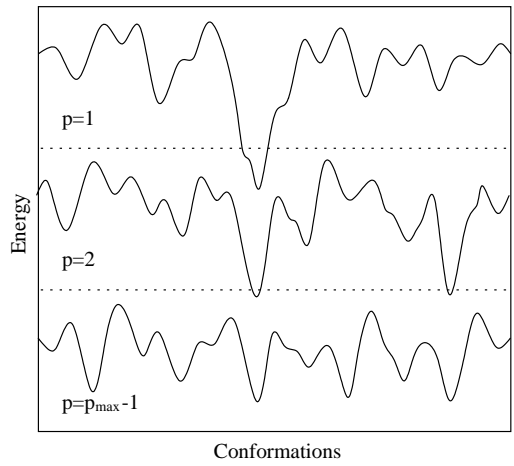}
    \vspace*{0.15in}
    \caption[]
    {Energy landscapes of sequences trained to be thermodynamically stable
    in a one, two and $p_{\rm max} - 1$ target conformations.  
    As the number of targets increases, the depth to which the target wells 
    can be trained diminishes.  At $p = p_{\rm max}$, the wells are 
    lost among nearby fluctuations.
    }
  \label{capacity_thermo}
  \end{center}
\end{figure}

What happens to the protein energy landscape upon introducing further
target conformations?  Consider an energy landscape in which there lies a
single well of maximal depth.
As a second (and, by assumption, independent) well
is introduced, the depth of the first is reduced
(Figure \ref{capacity_thermo}).  
As $p$ approaches $p_{\rm max}$, the typical well depth diminishes 
such that, at $p = p_{\rm max}$, the minima are indistinguishable from nearby 
fluctuations in the landscape.

For a uniform composition ({\it i.e.}, a homopolymer), zero conformations 
are encodable, as expected.  Frequently studied binary models allow at 
most one configuration to be stored, while for a 20 amino acid set, 
$p_{\rm max} \simeq 4.67$.
In all cases, as $p$ approaches $p_{\rm max}$, the minima become increasingly 
nominal.  It may be possible to find a binary ({\it e.g.}, H-P) 
sequence with a global minimum above an arbitrary compact target.
But there is typically of order one sequence per conformation, 
and the sequence is statistically unlikely to be stable. 
In this sense, binary models are not accurate representations of proteins.


{\bf Application to Heteropolymer Design and Prions}
Our results may be considered in the more general context of 
heteropolymer engineering and rational drug design.
The ability to remember multiple conformations admits a potentially 
dramatic increase in the variety of heteropolymer function.
We have provided arguments that training to superimposed contact maps provides
a viable method of designing multiply-conforming sequences.
To what extent can we exercise control over their occupied conformations? 

Shakhnovich and co-workers \cite{Shak98} observed in simulation what they refer
to as kinetic partitioning: some sequences designed to be 
stable in two conformations initially fold to one structure before 
later folding to the other.  On time scales short by comparison, 
the distribution over conformations occurs according to kinetic
accessibility rather than conformational stability.
We are investigating the extent to which temperature can be used to 
effect a change of the dominant occupied conformation before the onset of
equilibrium.

A naturally occurring and much studied heteropolymer thought to possess 
multiple stable conformations is prion protein \cite{QUATERNARY}.
Prions are infectious, transmissible pathogens
composed exclusively of the modified protein ${\rm PrP}^{\rm Sc}$ \cite{SP98}.
The chemical (primary) structure of ${\rm PrP}^{\rm Sc}$ is identical to the 
normal prion protein ${\rm PrP}^{\rm C}$, but its conformation 
(tertiary structure) is significantly different.
Prion diseases, such as BSE, CJD and scrapie of sheep, are believed to result
from the conformational conversion of 
${\rm PrP}^{\rm C}$ to ${\rm PrP}^{\rm Sc}$ and the resulting accumulation 
of the abnormal protein \cite{SP98}. 

Our calculations support the view that prion disease is caused by misfolding
to a second stable conformation.
Far from being confined to particular or correlated structures, 
the ability of a protein to take on multiple biologically active 
conformations is ubiquitous.
In addition to pathological proteins such as prions, 
we conjecture the existence of proteins which fold to multiple
biologically {\it useful} conformations.
Definitive observations to this end would have significant implications 
on our understanding of protein function.



\vspace*{-0.15in}

\begin{thebibliography}{99}
\bibitem{SCALING}
This assumes that the number of patterns stored does not scale more 
quickly than $N$, which we find below is reasonable.
\bibitem{CLT} This estimation is consistent with our use of the central limit
theorem provided $\sqrt{2} \sigma_{{\rm tot}_i} \sqrt{\ln A} < {z'p\over 2}
\sigma$, that is, $\ln A < {z'p\over 4}$.
\bibitem{INFO_U} The $A\times A$ pair potential $U$ is part of the
decoding apparatus 
(encoded by the laws of physics or included in the folding algorithm) 
and need not be transmitted.
\end{thebibliography}
\end{document}